# Amplification to Synthesis: A Comparative Analysis of Cognitive Operations Before and After Generative AI

Liz Cho

Dr. Dongwook Yoon

University of British Columbia

April 17, 2026

## Abstract

Cognitive operations are a rising concern in the geopolitical sphere, a quite yet rigorous fight for public perception and decision making. While such operations have been extensively studied in the context of bot-driven amplification, the emergence of generative AI introduces a new set of capabilities that may have fundamentally altered how these operations are designed and executed. The possible evolution of cognitive operation via generative AI puts nation states vulnerable without proper mitigation strategies. To address this, we compared behavioral and linguistic coordination patterns in X (formerly Twitter) datasets from the 2016 and 2024 U.S. presidential elections. Utilizing a combined corpus of over 133,000 posts, we applied post-type distribution, semantic clustering, temporal synchrony analysis, and Jaccard-based lexical overlap measures.

Findings suggest that the 2024 corpus exhibits a distinct pattern from 2016. Original content rose from 59% to 93% with retweets virtually disappeared; lexical overlap collapsed from a mean Jaccard score of 0.99 to 0.27, with posts converging on the same subject matter expressed in markedly different words; and temporal coordination shifted from pervasive cross-semantic synchrony to narratively concentrated co-occurrence. Taken together, these patterns point toward an operational logic organized around active content generation and narrative-specific targeting—characteristics consistent with generative AI involvement. These findings offer an empirical baseline for future research investigating generative AI's role in the cognitive operation pipeline, and as a practical reference point for security practitioners developing detection frameworks calibrated to the post-generative AI threat environment.

## Introduction

Cognitive warfare is an emerging military and academic concept that weaponizes information—through the strategic deployment of traditional and emerging technologies—to disrupt, undermine, influence, or modify human perception and decision making (Backes & Swab, 2026; Deppe & Schaal, 2024; Wirges, 2026). Evidence for cognitive operations is increasingly detected across various geopolitical spheres. For example, 2016 and 2024 U.S. presidential elections was heavily reported to be targeted by foreign interference (Deppe & Schaal, 2024), alongside deliberate disinformation campaigns regarding the origins of COVID-19 and the Ukraine Euromaidan crisis (Ebbott et al., 2021; Sedova et al., 2021).

Nation-states promptly recognized cognitive operations as a primary vector of geopolitical conflict, framing as “mind as battle field” (Defence, 2021) and the “sixth domain of warfare (Le Guyader, 2022). Accordingly, understanding the underlying operational architecture has become critical for developing robust mitigation strategies (Deppe, 2023; Romanishyn et al., 2025; Sedova et al., 2021).

In this context, generative AI was investigated as a key technology in the contemporary cognitive operation (Schroeder et al., 2025). In particular, the literature highlights the capacity of Large Language Models (LLMs) to generate human-like, persuasive narratives at scale—capabilities that may be exploited by malicious actors. These concerns call for empirical evidence on how LLMs may have altered the operational patterns, such that national security communities can identify potential gaps in the emerging tactics to develop robust cognitive defense frameworks.

Accordingly, this paper investigates following research questions:

- **Primary RQ:** Do cognitive operations in the post–generative-AI era exhibit distinct patterns that were not observed prior to the emergence of generative-AI, and that correspond to characteristics associated with AI-assisted content generation?
- **Sub RQ1:** How do behavioural coordination patterns in political messaging differ between the 2016 and 2024 datasets?

- **Sub RQ2**: How do linguistic patterns differ between coordinated messaging in the 2016 and 2024 datasets?

By empirically examining shifts in coordination dynamics and linguistic generation patterns, this study offers empirical grounding for policymakers and security practitioners to develop informed mitigation strategies in response to the evolving landscape of cognitive operations.

## Background

A robust body of literature established the early foundational contours of cognitive operations. One pillar of literature studied X (formally Twitter) activity of Russia's Internet Research Agency (IRA) on the U.S. 2016 presidential election to identify its structural patterns. Badawy et al. (2018) found conservative accounts produced 3.2 times more tweets than liberal accounts, with this content further amplified by other conservative users which included a mix of organic participants and right-wing trolls. Similar findings were made by Golovchenko et al. (2020) where conservative trolls were more active than liberal ones. Linvill and Warren (2020b, 2020a) further investigated the IRA's behavioural pattern, characterizing IRA accounts across five functional categories and tracing a three-phase lifecycle—introduction, growth, and amplification—through which the network strategically cultivated public engagement over the election cycle. They together reveal the mix of retweets and bot networks for systematic exploitation of desired narrative.

A parallel body of literature moved beyond individual case studies to identify the enabling technologies of large-scale information operations. Prior to the emergence of generative AI, automated bot networks served as the primary infrastructure for narrative amplification. Shao et al. (2018) offer early empirical grounding for this claim. The study conducted a longitudinal analysis of X (formally Twitter) data from 2016 to 2017 and found social bots amplified low-credibility contents. Building on such findings, Sedova et al. (2021)

formalized this mechanism into an operational framework, positioning bot networks as a central technological vector that disseminated disinformation.

Following the commercial proliferation of LLMs in late 2022, a novel body of research emerged. This body of research documented the capacity of these models to generate highly personalized persuasive messaging at scale. Matz et al. (2024) provide early empirical grounding for this concern, demonstrating that politically oriented messages crafted by ChatGPT exhibited significantly greater persuasive influence than non-personalized equivalents. Simchon et al. (2024) found that LLM-generated political advertisements tailored to recipients' personality profiles were measurably more effective than generic counterparts. Extending this line of evidence, Hackenburg and Margetts (2024) showed that behavioral data of users could be directly piped into LLMs to produce individualized political persuasion content at scale. Collectively, these findings have led researchers to raise the possibility of deliberate weaponization of LLMs as part of the Cognitive Operation pipeline — that would substantially expand the velocity, variety, and volume of disinformation (Barman et al., 2024).

## Method

### Study Objective

The study objective is to identify systematic structural shifts in coordinated influence operations following the advent of generative AI. In particular, we aim to understand how three operational dimensions — post-type distribution, temporal synchrony, and linguistic similarity — differ across the two election cycles and whether those differences are consistent with the introduction of generative AI as an enabling technology.

### Data Acquisition

The study draws on two publicly available tweet dataset corresponding to the 2016 and 2024 U.S. presidential election cycle. The 2016 dataset was sourced from the FiveThirtyEight Russian Troll Tweets repository (FiveThirtyEight, 2018/2018), a widely cited corpus formally attributed to a state-linked influence operation reported by X (Linvill & Warren, 2020b). The 2024 dataset was obtained from the GitHub repository maintained by a research group at the University of Southern California, which compiled 2024 U.S election-related tweets.

To enable meaningful cross-temporal comparison, the analysis focused on tweets within the fixed timewindow—2016-09-26-16:50 to 2016-09-26-23:59 and 2024-09-10-16:50 to 2024-09-10-23:59—corresponding to a single defining event within each election cycle—the presidential debate. Each yielded $n$ = 1063 and $n$ = 132,923 for 2016 and 2024, respectively.

## Measures

### *Post-Type Distribution*

To unpack the possible difference in coordination pattern (sub R1), we first examined post-type distribution across the two election datasets. We built on the premise that a generative-AI-augmented operation would produce a measurably higher proportion of original posts, reflecting the technology's capacity to fabricate novel content at scale. Each tweet was classified into one of three mutually exclusive categories — original post, retweet, or quote tweet — and their percentage distribution was compared between the 2016 and 2024 corpora.

### *Temporal Synchrony*

Considering that burst-like coordination in posting activity is widely recognized as a behavioral indicator of inorganic, orchestrated activity (Ebbott et al., 2021), we then examined temporal synchrony patterns across the two corpora. Given generative AI's capacity to produce semantically targeted narratives, temporal synchrony was analyzed under two conditions: 1) within-semantic cluster and 2) cross-semantic cluster. We aimed to assess whether post generative-AI campaigns exhibit not merely elevated coordination, but increasingly targeted coordination concentrated within thematically coherent content groups.

To address the inconsistent volume between 2016 and 2024 datasets, we randomly selected 1,063 posts from the 2024 dataset to match the total volume of the 2016 corpus. This ensured that observed differences in temporal synchrony reflect behavioral patterns rather than artifacts of corpus size.

## Lexical Similarity

Addressing Sub RQ2, we calculated pairwise lexical overlap scores within each semantic cluster. We employed Jaccard similarity coefficient—measures how similar two sets are based on the elements they share—yielding a score range from 0 (no overlap) to 1 (identical lexical content). By comparing the Jaccard scores across the two election cycles, we assessed whether post-generative AI campaigns exhibit higher degrees of lexical heterogeneity.

## Semantic Clustering Procedure

### *Preprocessing*

The corpus was first imported into a Python environment using the Pandas data analysis library. To reduce noise and eliminate non-semantic artifacts common in social media text, a preprocessing pipeline was applied to each tweet. Specifically, URLs and user mentions were removed using regular expression filters, and newline characters were replaced with whitespace to normalize textual formatting. Tweets with fewer than 20 characters were further excluded, as very short posts often consist of isolated hashtags, links, or truncated fragments that provide insufficient semantic information for embedding-based analysis. For lexical analysis, non-English tweets were further excluded, as the Jaccard similarity coefficient requires comparison between texts in the same language.

This process resulted in four preprocessed datasets: temporal coordination (2016, $n$ = 1,063; 2024, $n$ = 132,923) and lexical coordination (2016, $n$ = 980; 2024, $n$ = 118,896).

***Sentence Embedding Representation***

To capture semantic similarity across tweets, each preprocessed tweet was transformed into a dense vector representation using the Sentence-BERT framework. Considering the multilingual nature of the temporal coordination dataset, a multilingual model was employed, named paraphrase-multilingual-MiniLM-L12-v2 (Reimers & Gurevych, 2019). For lexical similarity analysis, all-MiniLM-L6-v2, from the Sentence-Transformers library (Reimers & Gurevych, 2019) was employed.

***Clustering Algorithm***

Clusters of semantically related tweets were identified using the community detection algorithm implemented in the sentence-transformers utility library. This method constructs clusters by identifying groups of embeddings whose pairwise cosine similarity exceeds a specified threshold.

Cluster formation was governed by two key parameters: similarity threshold and minimum community size.

- **Similarity threshold.** The similarity threshold defines the minimum degree of semantic proximity required for two tweets to be linked together. Tweets were grouped into the same cluster only if their cosine similarity exceeded 0.82, ensuring high semantic coherence while allowing for paraphrased variations typical of coordinated messaging.
- **Minimum community size.** The minimum community size defines the smallest number of nodes required for a cluster to be considered analytically meaningful. This parameter was used to distinguish coordinated cognitive operations from organic user activity (e.g., purely coincidental topic convergence).

### *Parameter Calibration*

**2024 Dataset.** To prevent the arbitrary assignment of this parameter, a qualitative analysis was conducted by varying the minimum community size while holding the similarity threshold constant at 0.82. The objective was to identify the optimal density 'sweet spot' that successfully clusters semantically coherent narratives as tight as it can.

Initial attempts to establish this threshold relied on standard sub-linear scaling models using natural logarithm of the dataset volume ($\ln(N)$). This yielded threshold approximately 12 for both the temporal and lexical analyses datasets ($n$ = 132,923; $n$ = 118,896). While logarithmic scaling is frequently employed in data science to estimate baseline community density, this purely mathematical approach proved ineffective for both datasets. The resulting clusters lacked internal semantic coherence and failed to capture identifiable narrative structures. Therefore, the study manually inspected the semantic noise floor, identifying a boundary at 60 nodes for both datasets. Clusters below this threshold consistently lacked semantic coherence, whereas higher thresholds proved overly restrictive by excluding meaningful mid-sized clusters. Accordingly, a minimum community size of 60 was selected for the 2024 dataset across both temporal and lexical analyses.

**2016 Dataset.** A parallel calibration process was applied to the 2016 dataset. The ln(*N*) estimate yielded an initial minimum cluster size of approximately 7 for both temporal and lexical analyses (*n* = 1,063; *n* = 980). However, manual inspection showed that this threshold excluded several thematically coherent clusters that were smaller in size, but highly consistent, indicating that it was overly restrictive. The minimum cluster size was therefore gradually reduced from 7 to 5. At this level, meaningful coordinated narratives were retained while excluding clusters that are too small in size to reliably infer thematic consistency for both temporal and lexical dataset. Accordingly, a minimum community size of 5 was used for both analyses.

### *Final Cluster Output*

Put together, the process yielded 29 clusters (*n* = 282) and 109 clusters (*n* = 9,586) for the temporal synchrony analysis, and 29 clusters (*n* = 269) and 58 clusters (*n* = 8,040) for the lexical analysis, for the 2016 and 2024 datasets, respectively.

# Findings

## Post types shifted from retweet-driven amplification to original content generation

The 2016 dataset was characterized by substantial reliance on retweets (39.98%), with original posts comprising 59.08% and quoted tweets 0.94%. In contrast, the 2024 dataset was overwhelmingly dominated by original posts (93.19%), with quoted tweets increasing to 6.80% and retweets nearly absent (0.01%) (see Table 1). Because the quoted tweet function requires operators to append original commentary to existing posts, this increase aligns with the network's shift toward original content generation. Taken together, these findings suggest a structural shift in 2024 from retweet-driven amplification to the active synthesis of original, narrative-specific content.

**Table 1**

*Percentage Distribution of Post Typologies (2016 vs. 2024)*

| Post Type | 2016 | 2024 |
|---|---|---|
| retweets (%) | 39.98 | 0.01 |
| original (%) | 59.08 | 93.19 |
| quoted tweet (%) | 0.94 | 6.8 |

**Figure 1**

*Percentage Distribution of Post Typologies (2016 vs. 2024)*

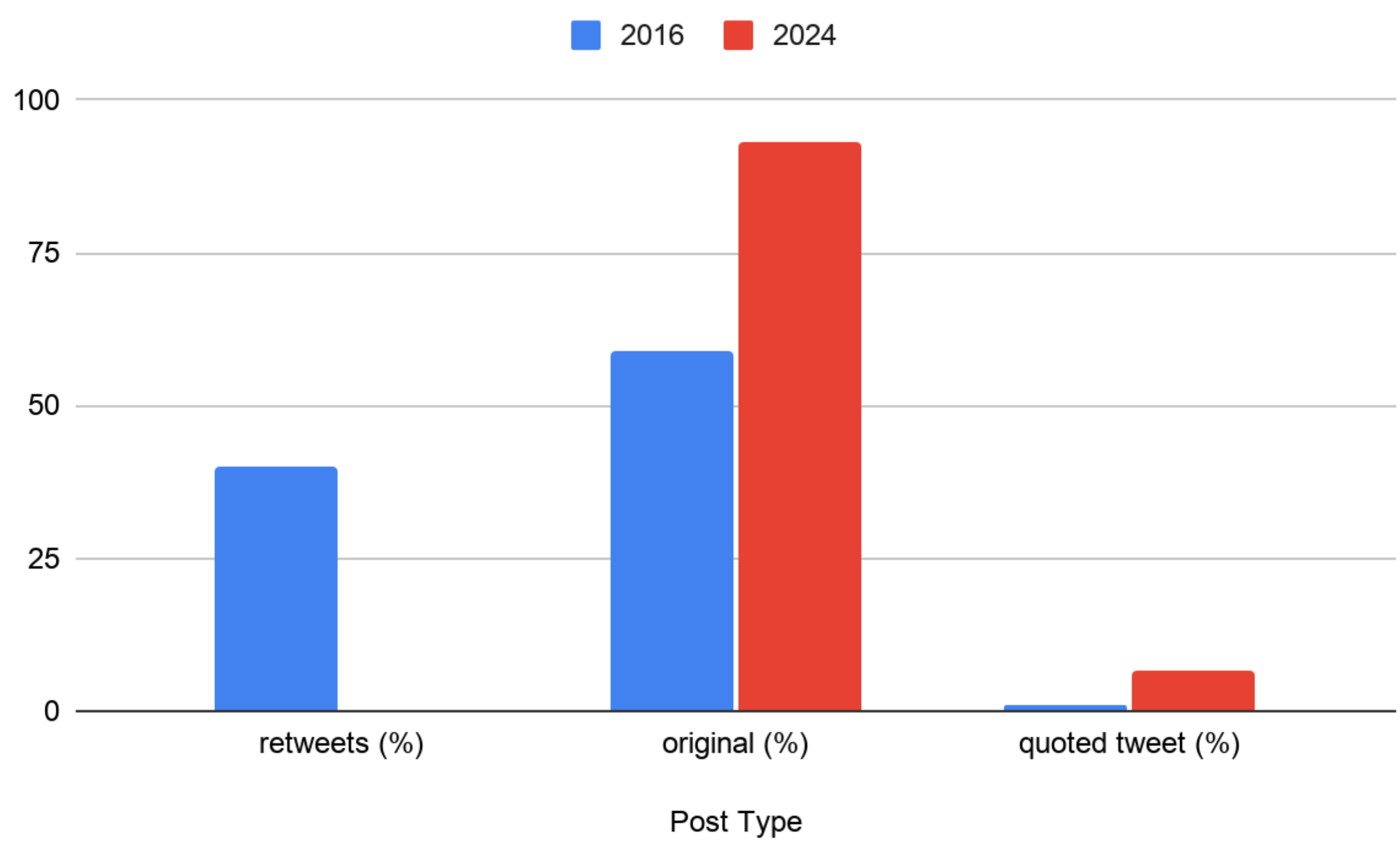

**Temporal coordination shifted from broad cross-thematic synchronization to concentrated, within-narrative coordination**

Under the cross-semantic condition (volume controlled at $n$ = 1,063), the 2016 dataset showed widespread temporal synchrony: 967 of 1,063 posts (91.0%) occurred in synchronized epochs (co-occurrence ≥ 2), and 238 of 334 epochs (71.3%) showed co-occurrence, with up to 16 posts appearing in the same second (see Table 2). In contrast, the 2024 dataset showed minimal synchrony: only 88 of 1,063 posts (8.3%) occurred in synchronized epochs, and 42 of 1,017 epochs (4.1%) showed co-occurrence, with a maximum of 3 simultaneous posts (see Table 3). The results suggest that temporal scynhronization within cross-thematic condition was common in 2016, but went nearly absent in 2024.

**Table 2**

*Distribution of Temporal Synchrony (2016)*

| Epoch Synchrony Level | Number of Epochs | Number of Posts Involved |
|---|---|---|
| 16 | 1 | 16 |
| 13 | 2 | 26 |
| 12 | 1 | 12 |
| 11 | 2 | 22 |
| 10 | 4 | 40 |
| 9 | 4 | 36 |
| 8 | 8 | 64 |
| 7 | 10 | 70 |
| 6 | 16 | 96 |
| 5 | 31 | 155 |

| 4 | 27 | 108 |
|---|---|---|
| 3 | 58 | 174 |
| 2 | 74 | 148 |
| 1 | 96 | 96 |

**Table 3**

*Distribution of Temporal Synchrony (2024)*

| Epoch Synchrony Level | Number of Epochs | Number of Posts Involved |
|---|---|---|
| 1 | 975 | 975 |
| 2 | 38 | 76 |
| 3 | 4 | 12 |

The pattern reverses under the within-cluster condition. The study examined two most temporally sycnhronized cluster in each dataset — Cluster #1 in 2016 and Cluster #3 in 2024 — effectively setting the ceiling for temporal sychrononization level. The 2016 corpus exhibited minimal within-cluster coordination, with at most 2 posts co-occurring simultaneously (see Table 4). The 2024 corpus, by contrast, demonstrated substantially higher within-cluster synchrony, with a peak of 26 simultaneous posts (see Table 5). Given the substantial volume disparity between the two clusters (2016: $n$ = 15; 2024: $n$ = 616), proportional metrics were applied for comparison. In 2016, 2 of 15 posts (13.3%) were involved in synchronized epochs, with 1 of 14 epochs (7.1%) exhibiting co-occurrence. In 2024, 385 of 616 posts (62.5%) were involved in synchronized epochs, with 80 of 311 epochs (25.7%) exhibiting co-occurrence (see Table 6).

**Table 4**

*Distribution of Temporal Synchrony within Cluster #1 (2016)*

| Epoch Synchrony Level | Number of Epochs | Number of Posts Involved |
|---|---|---|
| 1 | 13 | 13 |
| 2 | 1 | 2 |

**Table 5**

*Distribution of Temporal Synchrony within Cluster #3 (2024)*

| Epoch Synchrony Level | Number of Epochs | Number of Posts Involved |
|---|---|---|
| 1 | 231 | 231 |
| 2 | 41 | 82 |
| 3 | 16 | 48 |
| 4 | 3 | 12 |
| 5 | 3 | 15 |
| 6 | 3 | 18 |
| 7 | 2 | 14 |
| 8 | 1 | 8 |
| 12 | 1 | 12 |
| 13 | 1 | 13 |
| 14 | 2 | 28 |
| 15 | 1 | 15 |
| 17 | 1 | 17 |

| 18 | 2 | 36 |
|---|---|---|
| 19 | 1 | 19 |
| 22 | 1 | 22 |
| 26 | 1 | 26 |

**Table 6**

*Within-Cluster Temporal Synchrony (2016 vs 2024)*

| Year | Posts in Synchronized Epochs (%) | Epochs Exhibiting Co-occurrence (%) |
|---|---|---|
| 2016 | 13.3 | 7.1 |
| 2024 | 62.5 | 25.7 |

**Figure 2**

*Within-Cluster Temporal Synchrony (2016 vs 2024)*

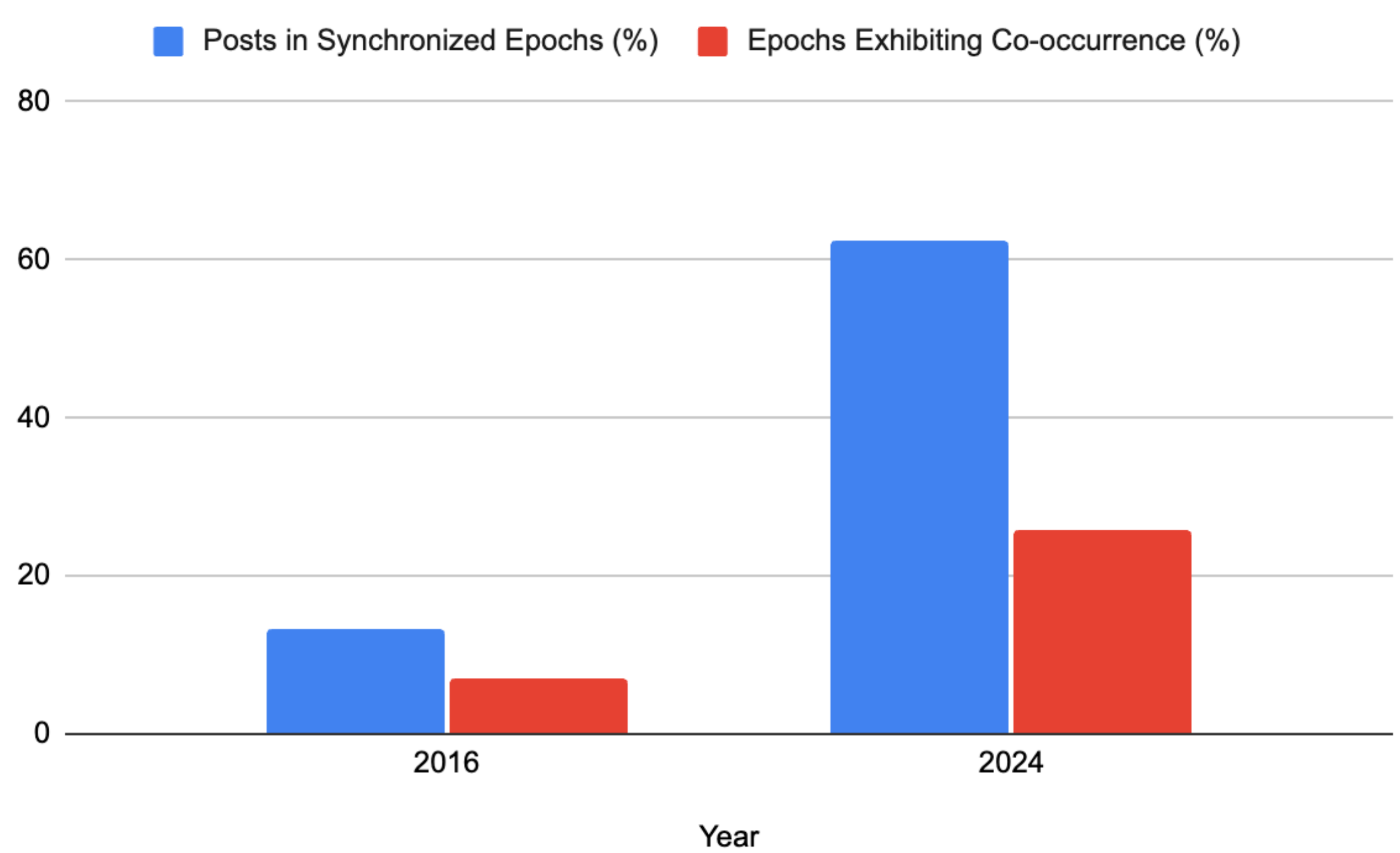

The two conditions produce a clean empirical dissociation: coordination in 2016 was pervasive across unrelated contents but nearly absent within the coherent narratives, while coordination in 2024 was concentrated within thematically coherent clusters. This inversion indicates a structural shift in 2024 from diffuse, cross-thematic amplification toward highly concentrated coordination within thematically coherent narratives.

**Linguistic patterns shifted from near-identical repetition in 2016 to diverse narrative expression in 2024**

The 2016 dataset exhibited extremely high lexical uniformity across clusters. Specifically, 26 of 28 clusters showed a lexical overlap score of 1.0 (mean = 0.99, median = 1.00, SD = 0.102, $n$ = 269), indicating exact duplication of content. For example, Cluster 1 consisted 15 unique posts containing the identical text: “Is this a public library or a typhoid ward? I'm wearing a Hazmat suit tomorrow!”, each published by different accounts. Upon inspection, these posts appeared as retweets, a pattern consistent with the earlier post typology analysis, which showed that the 2016 dataset was heavily dominated by retweet activity. A small number of exceptions appeared in Clusters 5 and 28, where one post exhibited a lexical overlap score of 0.238 and four posts exhibited a score of 0.25. However, upon manual inspection, the underlying narrative structure remained largely unchanged, with only minor lexical variations observed (see Table 7 and 8).

**Table 7**

*Example of Lexical Overlap within Cluster 5 in the 2016 English-Only Dataset*

| Representative node | one post with lexical overlap score = 0.238 |
|---|---|
| #ThereIsMoreThanOne way to skin a cat...but they will all piss it off. | #ThereIsMoreThanOne Way to Skin a Cat is the most horrific saying I've ever heard. |

**Table 8**

*Example of Lexical Overlap within Cluster 28 in the 2016 English-Only Dataset*

| Representative node | four posts with lexical overlap score = 0.25 |
|---|---|
| Breaking: Donald Trump invites Mark Geist, Benghazi survivor to debate! Great move! #debatenight #debates #debates2016 | BREAKING� Benghazi survivor Mark Geist confirmed to be on Trump's guest list tonight. #Debates2016 #debatenight |

In contrast, the 2024 dataset exhibited substantially lower lexical overlap (mean = 0.27, median = 0.23, SD = 0.167, *n* = 8,040), indicating considerably greater linguistic variation within clusters. For instance, Cluster #1 had a mean Jaccard score of 0.203 (median = 0.196, SD = 0.075), reflecting consistently low lexical overlap. As seen in Table 9, individual posts had different narrative frames while retaining the common theme — Taylor Swift's endorsement of Kamala Harris. Upon inspection, 98.18% (*n* = 1,347) appeared as the original post with 1.82% (*n* = 25) as quoted tweets and no retweets. This observation further aligns with the earlier post typology analysis, which showed that the 2024 dataset was dominated by original contents.

Together, the lexical analysis revealed linguistic structure shift across the two election cycles: the 2016 corpus was characterized by near-uniform lexical score consistent with retweet reproduction, while the 2024 corpus exhibited consistently low overlap across semantically coherent clusters, consistent with original, generatively produced content.

**Table 9**

*Example of Lexical Overlap within Cluster 1 in the 2024 English-Only Dataset*

| Post Content | Jaccard Similarity Score |
|---|---|
| Taylor Swift has formally endorsed Kamala Harris after the candidate’s debate with Donald Trump, following up the ABC-hosted match-up with a strong message pos... | 1 |
| Taylor Swift has formally endorsed Kamala Harris following the candidate’s debate with Donald Trump, following the debate on ABC with a strong message posted t... | 0.6538 |
| Taylor Swift has formally endorsed Kamala Harris after the candidate’s debate with Donald Trump, following up the ABC-hosted match-up with a strong message posted to her Instagram | 0.8077 |
| Taylor Swift has endorsed Kamala Harris in the presidential election following Tuesday night's debate, and also addressed Donald Trump using AI images of her. | 0.2432 |
| Superstar Taylor Swift has endorsed Kamala Harris in the US presidential race just moments after the Democrat nominee’s first debate with her Republican opponent Donald Trump wrapped up. #7NEWS | 0.2821 |
| Taylor Swift has endorsed Kamala Harris minutes after the vice president's fiery debate with Donald Trump. | 0.4074 |
| Taylor Swift has endorsed Kamala Harris for president | 0.3125 |

| just moments after the end of her presidential debate against Donald Trump. | |
|---|---|

## Discussion

This study examined whether post-generative AI cognitive operations exhibit behavioral and linguistic signatures distinct from those observed prior to the emergence of generative AI. Although a growing body of research has investigated coordinated inauthentic behavior through temporal synchrony and bot detection frameworks (Ebbott et al., 2021; Padalko, 2024; Schroeder et al., 2025; Stockwell et al., 2024), little empirical work has examined how generative AI may have structurally altered the operational signatures. The current findings suggest that this technological evolution may have manifested coordination pattern shift across three key dimensions: post-type distribution, semantic structure, and temporal coordination.

### Structural Departure from the Volume Amplification to Novel Content Generation

In the 2016 baseline, retweets comprised a substantial volume of all network activity. This finding is consistent with prior literature that identifies mass retweeting as a foundational tactic of the Russian Internet Research Agency (IRA) (Bovet & Makse, 2019; Linvill & Warren, 2020b). By 2024, however, this legacy tactic had virtually disappeared. Retweets were nearly absent (0.01%), with the corpus dominated by original posts and quoted tweets (99%). This signals a transition in the underlying production logic: naive reproduction in 2016 to novel content generation in 2024.

If the 2024 corpus was dominated by originally produced content, then the question becomes: how similar was that content? The lexical analysis provides an answer. In 2016, posts within the same semantic cluster were nearly identical in wording — exactly what would be expected if most content was retweeted. In 2024, posts addressing the same topic were expressed in a variety of narrative frames. On the surface, this resembles the kind of variation one would expect from genuine, independent users. The concern, however, is that producing large volumes of topically consistent but lexically varied content is a well documented capability of LLMs. Notably, Matz et al. (2020) and Padalko (2024) describe that AI can systematically tailor the framing of the same message to different audiences. The 2024 lexical profile — high semantic coherence and low lexical similarity — is consistent with this kind of targeted, varied content generation, suggesting the possibile generative AI footprint.

**Narrative-Focused Temporal Coordination as an Emerging Operational Signature**

In 2016, cross-semantic temporal synchrony was pervasive. The majority of epochs exhibited co-occurrence, with the posts sharing unrelated contents. This pattern aligns closely with what the existing literature describes as the defining behavioral signature of bot-driven influence operations: automated networks functioning as quantity amplifiers, which is characterised by burst-like temporal synchrony (Ebbott et al., 2021). In 2024, this signature was nearly absent. Cross-semantic synchrony declined to a much less burst-like pattern, with a maximum of only three simultaneous posts. At first glance, this appears consistent with recent scholarship suggesting that coordinated actors adopt more heterogeneous posting rhythms to evade synchrony-based detection (Schroeder et al., 2025). However, the within-cluster analysis reveals the opposite pattern. In 2016, within-cluster synchrony was negligible, consisting of only a single co-occurring event. In contrast, the 2024 dataset exhibited substantially higher synchrony within semantic clusters, with 62.5% of posts and 25.7% of epochs co-occurring.

This suggests that coordination in 2024 has not disappeared, but has instead become more targeted. Rather than operating across unrelated content streams, temporal coordination appears concentrated within specific narratives, indicating a shift from diffuse to narrative-focused coordination.

**Implications for Generative AI in Cognitive Operations**

Taken together, our study revealed that the 2024 dataset had a distinct behavioural and lexical pattern from the 2016. Spanning the periods before and after the public proliferation of generative AI, the data demonstrates a structural shift from indiscriminate volumetric amplification toward narratively targeted, lexically diversified content synthesis. These metrics are highly consistent with the documented capabilities of generative AI to engineer bespoke messaging at scale (Padalko, 2024), suggesting possible footprint in modern cognitive operations (OpenAI, 2024).

**Attribution and Interpretive Caution**

It must be emphasized, however, that this study does not confirm that the observed operational shifts were driven by the incorporation of generative AI into the influence operation pipeline. The analysis is purely exploratory in design, and the findings are better understood as establishing an empirical baseline than as resolving questions of attribution or causation. The primary contribution is to document a set of behavioral and linguistic signatures that are both empirically distinguishable from per-AI era operations and are analytically coherent with the generative AI capabilities— providing an empirical reference point for national security researchers and practitioners.

## Limitations

While this study establishes an exploratory structural baseline for analyzing generative AI-assisted cognitive operations, several methodological limitations bear on the interpretation of the findings and the generalizability of the conclusions.

### *Platform and Behavioral Confounders*

The eight-year gap introduces inherent environmental variables. Most notably, the typological shift toward original content generation observed in the 2024 corpus may partly reflect broader changes in organic user behavior on X (formally Twitter) over this period. Therefore, the extent to which the observed compositional shift reflects operational design rather than platform ecology remains an open question.

### *Data Collection and Sampling Bias*

Because the 2016 and 2024 corpora were compiled by different research teams, variances in their API extraction protocols and keyword filtering may introduce systematic sampling bias. These methodological discrepancies may systematically skew the comparative analysis, resulting in the artificial over- or underrepresentation of specific post typologies and temporal coordination clusters.

### *Temporal Scope and Replication*

The analysis is confined to a single, 7-hour operational window. Therefore, we cannot generalize the observed pattern in other temporal windows or to other geopolitical events. Future research must expand this methodology across diverse, longitudinal timeframes to verify whether the findings are generalizable across all modern cognitive operations.

### *Modality Constraints*

The analysis is restricted to textual content and does not account for the multimodal capabilities of contemporary generative AI, including synthetic audio, deepfake imagery, and AI-generated video. A growing body of documented cases suggests that visual and audiovisual synthetic content may serve to reinforce or amplify textual narratives, lending them added credibility and affective resonance (Groh et al., 2024; Hameleers et al., 2024). Restricting the analytical lens to text therefore risks underestimating the full operational scope of generative AI-assisted influence campaigns. Future research should integrate multimodal analysis to construct a more comprehensive account of how operational patterns have changed across the pre- and post-generative AI.

### *Platform Constraints*

The study is further confined to the X (formally Twitter) ecosystem, where text is a primary information exchange modality. Coordinated influence operations have been documented across a diverse range of platforms including Meta and TikTok (Dommett & Power, 2019; Park et al., 2023), whose architectures are organized primarily around images and video rather than text. The behavioral and linguistic signatures identified here may not translate directly to these environments, and the platform-specific nature of the findings limits the generalizability. Therefore, cross-platform comparison is necessary for a more comprehensive understanding of post-generative AI operational change.

### *Unexplored Signatures*

While this study identifies critical shifts in typological and lexical patterns, it does not exhaustively map all theoretically relevant signatures of generative AI integration. Two dimensions warrant particular attention in future empirical work. First, linguistic diversity—the capacity of LLM to translate multilingual content. Second, temporal latency—the speed at which an automated network synthesizes and deploys novel content in response to real-world stimuli. Expanding future analytical frameworks to incorporate these dimensions will significantly enrich both threat detection architectures and the theoretical understanding of generative AI's operational deployment.

**References**

Backes, O., & Swab, A. (2026, March 27). *Cognitive Warfare: The Russian Threat to Election Integrity in the Baltic States | The Belfer Center for Science and International Affairs*. https://www.belfercenter.org/publication/cognitive-warfare-russian-threat-election-integrity-baltic-states

Badawy, A., Addawood, A., Lerman, K., & Ferrara, E. (2018). *Characterizing the 2016 Russian IRA Influence Campaign* (arXiv:1812.01997). arXiv. https://doi.org/10.48550/arXiv.1812.01997

Bakirov, A., & Suleimenov, I. (2025). Theoretical Bases of Methods of Counteraction to Modern Forms of Information Warfare. *Computers*, *14*(10), 410. https://doi.org/10.3390/computers14100410

Barman, D., Guo, Z., & Conlan, O. (2024). The Dark Side of Language Models: Exploring the Potential of LLMs in Multimedia Disinformation Generation and Dissemination. *Machine Learning with Applications*, *16*, 100545. https://doi.org/10.1016/j.mlwa.2024.100545

Bovet, A., & Makse, H. A. (2019). Influence of fake news in Twitter during the 2016 US presidential election. *Nature Communications*, *10*(1), 7. https://doi.org/10.1038/s41467-018-07761-2

Defence, N. (2021, September 10). *Fall 2021 NATO Innovation Challenge*. https://www.canada.ca/en/department-national-defence/campaigns/fall-2021-nato-innovation-challenge.html

Deppe, C. (2023). *Disinformation In Cognitive Warfare, Foreign Information Manipulation And Interference, And Hybrid Threats*. https://doi.org/10.5281/zenodo.10005172

Deppe, C., & Schaal, G. S. (2024). Cognitive warfare: A conceptual analysis of the NATO ACT cognitive warfare exploratory concept. *Frontiers in Big Data*, *7*, 1452129. https://doi.org/10.3389/fdata.2024.1452129

Ebbott, E., Saletta, M., & Stearne, R. (2021). *Understanding Mass Influence—A case study of the Internet Research Agency as a contemporary mass influence operation* (Understanding Mass Influence). The University of Melbourne, University of New South Wales, The University of Adelaide, Edith Cowan University, and Macquarie University.

FiveThirtyEight. (2018). *Fivethirtyeight/russian-troll-tweets* [Computer software]. https://github.com/fivethirtyeight/russian-troll-tweets (Original work published 2018)

Golovchenko, Y., Buntain, C., Eady, G., Brown, M. A., & Tucker, J. A. (2020). Cross-Platform State Propaganda: Russian Trolls on Twitter and YouTube during the 2016 U.S. Presidential Election. *The International Journal of Press/Politics*, *25*(3), 357–389. https://doi.org/10.1177/1940161220912682

Groh, M., Sankaranarayanan, A., Kim, D., Lippman, A., & Picard, R. (2024). Human detection of political speech deepfakes across transcripts, audio, and video. *Nature Communications*, *15*. https://doi.org/10.1038/s41467-024-51998-z

Hackenburg, K., & Margetts, H. (2024). Evaluating the persuasive influence of political microtargeting with large language models. *Proceedings of the National Academy of Sciences of the United States of America*, *121*(24), e2403116121. https://doi.org/10.1073/pnas.2403116121

Hameleers, M., van der Meer, T. G. L. A., & Dobber, T. (2024). Distorting the truth versus blatant lies: The effects of different degrees of deception in domestic and foreign political deepfakes. *Computers in Human Behavior*, *152*, 108096. https://doi.org/10.1016/j.chb.2023.108096

Linvill, D. L., & Warren, P. L. (2020a). Engaging with others: How the IRA coordinated information operation made friends. *Harvard Kennedy School Misinformation Review*, *1*(2). https://doi.org/10.37016/mr-2020-011

Linvill, D. L., & Warren, P. L. (2020b). Troll Factories: Manufacturing Specialized Disinformation on Twitter. *Political Communication*, *37*(4), 447–467. https://doi.org/10.1080/10584609.2020.1718257

Matz, S. C., Teeny, J. D., Vaid, S. S., Peters, H., Harari, G. M., & Cerf, M. (2024). The potential of generative AI for personalized persuasion at scale. *Scientific Reports*, *14*(1), 4692. https://doi.org/10.1038/s41598-024-53755-0

Montgomery, M. L., RADM (Ret ). Mark. (2024, September 26). How U.S. Adversaries Undermine the Perception of Election Integrity. *FDD*. https://www.fdd.org/analysis/2024/09/26/how-u-s-adversaries-undermine-the-perception-of-election-integrity/

OpenAI. (2024). *Disrupting deceptive uses of AI by covert influence operations*. OpenAI. https://openai.com/index/disrupting-deceptive-uses-of-ai-by-covert-influence-operations/

Padalko, H. (2024). *AI and Information Manipulation: Russia's Interference in the US Elections*.

Reimers, N., & Gurevych, I. (2019). *Sentence-BERT: Sentence Embeddings using Siamese BERT-Networks* (arXiv:1908.10084). arXiv. https://doi.org/10.48550/arXiv.1908.10084

Romanishyn, A., Malytska, O., & Goncharuk, V. (2025). AI-driven disinformation: Policy recommendations for democratic resilience. *Frontiers in Artificial Intelligence*, *8*, 1569115. https://doi.org/10.3389/frai.2025.1569115

Schroeder, D. T., Cha, M., Baronchelli, A., Bostrom, N., Christakis, N. A., Garcia, D., Goldenberg, A., Kyrychenko, Y., Leyton-Brown, K., Lutz, N., Marcus, G., Menczer, F., Pennycook, G., Rand, D. G., Ressa, M., Schweitzer, F., Song, D., Summerfield, C., Tang, A., … Kunst, J. R. (2025). *How malicious AI swarms can threaten democracy: The fusion of agentic AI and LLMs marks a new frontier in information warfare* (Version 4). arXiv. https://doi.org/10.48550/ARXIV.2506.06299

Sedova, K., McNeill, C., Johnson, A., Joshi, A., & Wulkan, I. (2021). *AI and the Future of Disinformation Campaigns: Part 1: The RICHDATA Framework*. Center for Security and Emerging Technology. https://doi.org/10.51593/2021CA005

Shao, C., Ciampaglia, G. L., Varol, O., Yang, K.-C., Flammini, A., & Menczer, F. (2018). The spread of low-credibility content by social bots. *Nature Communications*, *9*(1), 4787. https://doi.org/10.1038/s41467-018-06930-7

Simchon, A., Edwards, M., & Lewandowsky, S. (2024). The persuasive effects of political microtargeting in the age of generative artificial intelligence. *PNAS Nexus*, *3*(2), pgae035. https://doi.org/10.1093/pnasnexus/pgae035

Stockwell, S., Hughes, M., & Swatton, P. (2024). *AI-Enabled Influence Operations: Safeguarding Future Elections*.

https://cetas.turing.ac.uk/publications/ai-enabled-influence-operations-safeguarding-future-elections

Wirges, J. (2026, March 18). Narrative as a Weapon: Russian, Iranian, and Chinese Approaches to Cognitive Warfare. *Small Wars Journal by Arizona State University*. https://smallwarsjournal.com/2026/03/18/narrative-as-a-weapon/